\documentclass[%
 reprint,
superscriptaddress,
 amsmath,
 amssymb,
 aps,
 prx,
 showpacs, 
 floatfix,
]{revtex4-2}

\usepackage{graphicx}% Include figure files
\usepackage{dcolumn}% Align table columns on decimal point
\usepackage{bm}% bold math
\usepackage{physics}% Dirac notations
\usepackage{hyperref}
\hypersetup{
  colorlinks   = true, %Colours links instead of ugly boxes
  urlcolor     = blue, %Colour for external hyperlinks
  linkcolor    = blue, %Colour of internal links
  citecolor   = blue %Colour of citations
}

\begin{document}

\preprint{APS/123-QED}
 
\title{Quantum-Acoustical Drude Peak Shift}% Force line breaks with \\

\author{J.~Keski-Rahkonen}
\email{joonas_keskirahkonen@fas.harvard.edu}
\affiliation{Department of Physics, Harvard University, Harvard University, Cambridge, MA 02138, USA}
\affiliation{Department of Chemistry and Chemical Biology, Harvard University, Cambridge,
MA 02138, USA}

\author{X.-Y.~Ouyang}
\affiliation{Yuanpei College, Peking University, No.5 Yiheyuan Rd, Beijing
100871, China}
\affiliation{School of Physics, Peking University, No.5 Yiheyuan Rd, Beijing
100871, China}

\author{S.~Yuan}
\affiliation{School of Physics, Peking University, No.5 Yiheyuan Rd, Beijing
100871, China}

\author{A.M.~Graf}
\affiliation{Harvard John A. Paulson School of Engineering and Applied Sciences,
Harvard, Cambridge, Massachusetts 02138, USA}

\author{A.~Aydin}
\affiliation{Department of Physics, Harvard University, Harvard University, Cambridge, MA 02138, USA}
\affiliation{Faculty of Engineering and Natural Sciences, Sabanci University, 34956 Tuzla, Istanbul, T{\"u}rkiye}

\author{E.J.~Heller}
\affiliation{Department of Physics, Harvard University, Harvard University, Cambridge, MA 02138, USA}
\affiliation{Department of Chemistry and Chemical Biology, Harvard University, Cambridge,
MA 02138, USA}


\date{\today}% It is always \today, today,

\begin{abstract}
\noindent

Quantum acoustics -- a recently developed framework parallel to quantum optics --  establishes a nonperturbative and coherent treatment of the electron-phonon interaction in real space. The quantum-acoustical representation reveals a displaced Drude peak hid  ing in plain sight within the venerable Fröhlich model: the optical conductivity exhibits a finite frequency maximum in the far-infrared range and the d.c. conductivity is suppressed. Our results elucidate the origin of the high-temperature absorption peaks in strange or bad metals, revealing that dynamical lattice disorder steers the system towards a non-Drude behavior.

\end{abstract}

\maketitle
%\tableofcontents

%\section{\label{sec:Introduction} Introduction}
\noindent
Stretching over four decades, an intensive theoretical pursuit has concentrated on finding an all-embracing explanation for a plethora of puzzling phenomena which have been colloquially labeled as ``bad'' or ``strange''. These kinds of ``bizarre'' materials seem to defy the traditional paradigms for electron behavior~\cite{rev.mod.phys_92_031001_2020} in metals. Mysteries abound,  such as high-temperature superconductivity beyond the grasp of the BCS theory~\cite{Rev.Mod.Phys_95_021001_2023, nat.rev.phys_3_462_2021}, the paradoxical existence of pseudogaps~\cite{science_314_1888_2006, rep.prog.phys_62_61_1999, Rev.Mod.Phys_75_473_2003, Rev.Mod.Phys_77_721_2005, Rev.Mod.Phys_79_353_2007} and charge density waves~\cite{Rev.Mod.Phys_60_1129_1988, nature_477_191_2011, nat.phys_8_871_2012, science_337_6096_2012}, the violation of the Mott-Ioffle-Regel (MIR) limit~\cite{Philos.Mag_84_2847_2004, Rev.Mod.Phys_75_1085_2003}, not to mention the major dilemma of linear-in-temperature resistivity over a wide temperature range (see, e.g., Refs.~\cite{Philos_Trans_R_Soc_A_369_1626_2011, science_336_6088_1554_2012, science_339_804_2013, nat.phys_15_142_2018, nat.phys_15_1011_2019, nature_595_667_2021, Phys.Rev.B_106_035107_2022}) at the mysterious but ubiquitous Planckian bound~\cite{rev.mod.phys_94_041002_2022}. This list of theoretical challenges also includes the elusive emergence of displaced Drude peaks (DDP)~\cite{Phys.Rev.Lett_81_2498_1998, Phys.Rev.B_66_041104_2002, Phys.Rev.Lett_93_167003_2004, Phys.Rev.B_55_14152_1997, Phys.Rev.B_68_134501_2003, Phys.Rev.Lett_82_1313_1999, Phys.Rev.Lett_75_105_1995, Phys.Rev.Lett_99_167402_2007, Phys.Rev.B_60_13011_1999, Phys.Rev.B_65_184436_2002, Phys.Rev.B_60_4342_1999, Phys.Rev.Lett_95_227801_2005, Phys.Rev.Lett_93_237007_2004, nat.phys_10_304_2014, j.condens.matter.phys_19-125208_2007}: a prominent absorption peak located typically in the infrared range, signaling a breakdown of the conventional Drude picture. This Letter reveals the origin of this phenomenon as the result of strong electron-phonon interaction if treated correctly as nonperturbative and coherent.

\begin{figure}[t]
\centering
\includegraphics[width=0.5\textwidth]{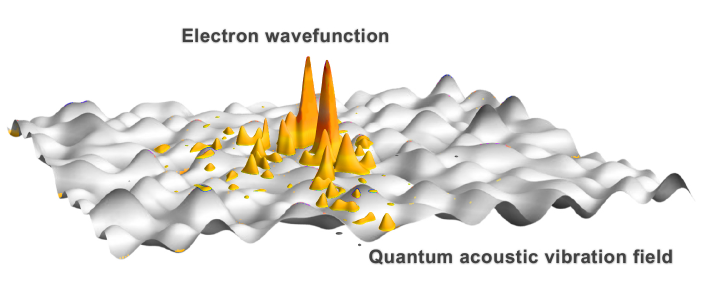}
\caption{Quantum acoustics. Illustration of the coherent state lattice vibrations at a certain temperature. Electrons experience a spatially continuous internal field formed by the thermal acoustic distortions. While undergoing quasi-elastic scattering events, like due to impurities, electrons can also be incipiently trapped by valleys of slowly undulating and propagating deformation potential when their kinetic energy is comparable to the fluctuations of the deformation potential.}
\label{Fig:DPeak_main}
\end{figure}

A putative culprit for the observed Drude shift in the optical conductivity peak has been suggested to be some as yet unidentified  dynamical disorder that generates evanescent localization, thus hampering but not precluding charge carrier diffusion~\cite{Rev.Mod.Phys_57_287_1985, Phys.Rev.B_64_155106_2001, j.phys.c_15_L707_1982, scipost.phys_11_39_2021, nat.commun_12_1571_2021}; Consequently, the zero-frequency conductivity does not vanish completely, but it is strongly suppressed, favoring the DDP phenomenology. This scenario contrasts with alternative points of view, like the common arguments resorting to collective modes~\cite{Phys.Rev.Lett_88_147001_2002, SciPost.Phys_3_25_2017} or to strong electron-electron correlations~\cite{Rev.Mod.Phys_83_471_2011,Lanzara2001}.

Here we show that a morphing potential landscape of hills and valleys stemming from thermal lattice vibrations by the Fr\"ohlich Hamiltonian (see Ref.~\cite{Heller22}), as illustrated in Fig.~\ref{Fig:DPeak_main} is the  sought-after sea of ``slowly moving bosonic impurities''~\cite{scipost.phys_11_39_2021}, or the cryptic ``self-induced randomness''~\cite{nat.commun_12_1571_2021, science_381_790_2023}. In a broader milieu, the subtle interplay between the Anderson localization and lattice vibrations has been encountered in a wide class of random metal alloys and other degenerate disordered systems. In fact, the intricate game of being localized or not was identified early on by Gogolin et al.~\cite{gogolin1, gogolin2} and Thouless~\cite{Phys.Rev.Lett_39_1167_1977}, even pondered by Anderson himself~\cite{Anderson_localization, Phys.Rev_109_1492_1958}. 

The random fluctuations introduced by lattice motion slowly but surely  scramble the quantum interference required for localization of the electronic state, resulting in \emph{transient} localization (for capturing the essential aspects of this phenomenon, see, e.g., Refs.~\cite{Phys.Rev.Lett_118_036602_2017}), which has lately been of interest in the context of crystalline organic semiconductors~\cite{Adv.Funct.Mater_26_2292_2016,Phys.Rev.Lett_96_086601_2006} and halide perovskites~\cite{Phys.Rev.Lett_124_196601_2020}. 

The quantum-acoustic route to linear and universal resistivity in strange metals~\cite{nature_manuscript} has opened up a new path unrelated to quantum criticality~\cite{sachdev1999quantum, Rev.Mod.Phy_94_035004_2022}, and not relying on (strong) electron-electron interaction~\cite{science_381_790_2023, sachdev2023quantum}, instead starting with the standard Fr\"olich Hamiltonian. 

Following the path paved in Refs.~\cite{Heller22, nature_manuscript}, here we demonstrate the formation of a DDP due to the electrons interacting with fluctuating lattice degrees of freedom. We go further by showing that this mechanism gives rise to a temperature dependence of spectral features in agreement with experimental DDP observations in strange metals. 

%\section{Theory}

More specifically, we consider the following Fr\"ohlich Hamiltonian~\cite{Frochlich1, Frochlich2} describing the lowest-order (linear) lattice-electron coupling~\cite{Mahan}:
\begin{equation}\label{Hamiltonian} 
\mathcal{H}_{\textrm{F}} = \sum_{\mathbf{p}}\varepsilon_{\mathbf{p}} c_{\mathbf{p}}c_\mathbf{p}^{\dagger} + \sum_{\mathbf{q}} \hbar \omega_{\mathbf{q}} a_{\mathbf{q}}^{\dagger} a_{\mathbf{q}} + \sum_{\mathbf{p}\mathbf{q}} g_{\mathbf{q}} c_{\mathbf{p} + \mathbf{q}}^{\dagger} c_{\mathbf{p}} \Big(a_{\mathbf{q}} + a_{\mathbf{-q}}^{\dagger} \Big)
\end{equation}
where $c_{\mathbf{p}}$ $(c_\mathbf{p}^{\dagger})$  is the creation (annihilation) operator for electrons with momentum $\mathbf{p}$ and energy $\varepsilon_{\mathbf{p}}$; whereas $a_{\mathbf{q}}$ $(a_\mathbf{q}^{\dagger})$ is the creation (annihilation) operator for longitudinal acoustic phonons of wave vector $\mathbf{q}$ and energy $\hbar \omega_{\mathbf{q}}$, respectively. The electron-phonon interaction is defined by its Fourier components $g_{\mathbf{q}}$. By following the steps of the recently established coherent state formalism in Ref.~\cite{Heller22,}, the Hamiltonian gives rise to an undulating and propagating potential landscape~\footnote{See Appendix~\ref{Appendix:def.potential} for showing a short derivation of the deformation potential from the Hamiltonian and explaining its key features; a more complete discussion on the coherent state formalism in the context of lattice vibrations and on the deformation potential can be found in Ref.~\cite{Heller22} and its relation to quantum transport in Ref.~\cite{nature_manuscript}.}: 
\begin{align}
    V_D(\mathbf{r},t)
    &=
    \sum_{\substack{\mathbf{q}}}^{\vert \mathbf{q} \vert \le q_D}
    g_{\mathbf{q}}
    \sqrt{\langle n_{\mathbf{q}}\rangle_{\textrm{th}}}
    \cos(\mathbf{q}\cdot\mathbf{r}-\omega_{\mathbf{q}}t+\varphi_{\mathbf{q}})
    \label{eq:VDcl}
\end{align}
where $q_D$ is Debye wavenumber, $\mathbf{r}$ is continuous position, $\varphi_{\mathbf{q}}=\textrm{arg}(\alpha_{\mathbf{q}})$ is the (random) phase of a coherent state $\vert \alpha_{\mathbf{q}} \rangle$, and  the mode population is determined by $\langle n_{\mathbf{q}}\rangle_{\textrm{th}}$. 

%The details behind deformation potential are further elucidated in Appendix~\ref{Appendix:def.potential}.  
 
The coherent state picture developed here is the dual partner of the traditional number state description of electron-lattice dynamics, so widely successful for describing electron resistivity~\cite{ashcroft1976solid}. In addition to recovering the results of the conventional Bloch-Gr{\"u}neisen theory~\cite{Bloch1930, Gruneisen1933other}, the coherent state representation extends beyond perturbation theory  (see Ref.~\cite{Heller22} for a more detailed discussion). The coherent state limit of quantum acoustics reveals a real-space, time-dependent description of electron-lattice interaction. A very similar notion was introduced in 1957 by Hanbury Brown and Twiss for the vector potential of a blackbody field~\cite{HBT1}, with the essential difference of missing the ultraviolet cut-off, i.e., the Debye wavenumber in the definition of our deformation potential originating from the minimal lattice spacing.

This follows the    quantum optics pathway pioneered  by Glauber~\cite{phys.rev_131_2766_1963}, a long neglected but essential wave perspective for lattice vibrations -- \emph{quantum acoustics}. Bardeen and Shockley, in the 1950's, regarded dynamical lattice distortions in nonpolar semiconductors~\cite{ShockleyBardeen1, ShockleyBardeen2}, and it seemed they would have been happy with a coherent state description, but the theory was subsumed by a number state, perturbative perspective.  
 
%\section{Results}

Within the present deformation potential framework, an electron undergoes quasielastic, coherence-preserving scattering events when roaming through the slowly altering potential landscape of hills and valleys. 

In this work, we focus on three prototypical compounds classified as strange/bad metals, namely LSCO, Bi2212 and Sr\textsubscript{3}Ru\textsubscript{2}O\textsubscript{7}~\footnote{See Appendix.~\ref{Appendix:material_parameters} for the material properties. Moreover, these parameters are in line with Ref.~\cite{nature_manuscript}.}. However, we want to stress that the physics we find below transcends the material-specific constraints: In general, dynamical disorder, caused by lattice vibrations here, temporarily confines electrons to nest in its instantaneous potential wells (see Fig.~\ref{Fig:DPeak_main}, a hallmark of transient localization dynamics,  resulting in the buildup of a DDP.    

%\subsection{Frozen approximation}

\begin{figure}[h!]
\centering
\includegraphics[width=.45\textwidth]{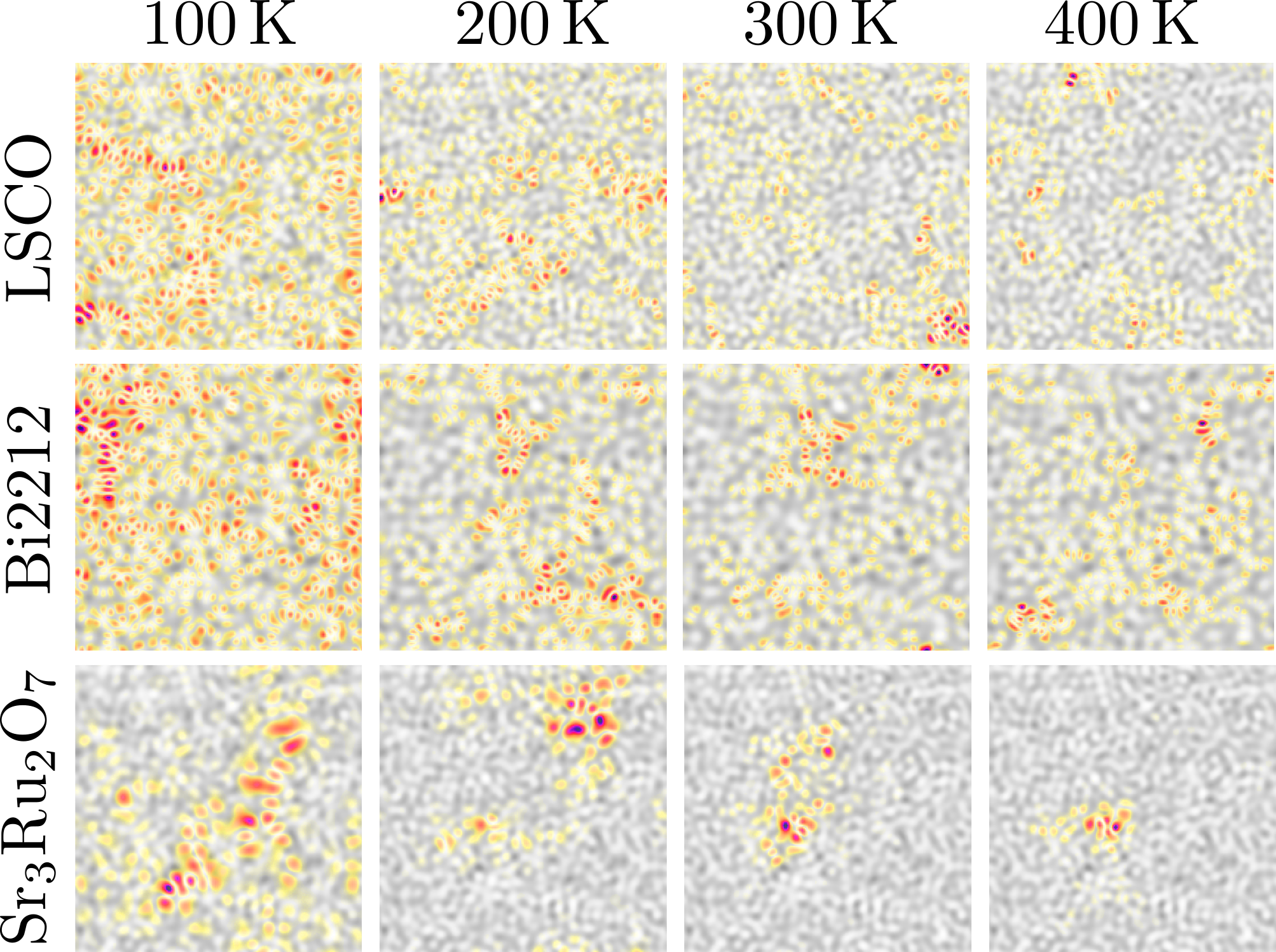}
\caption{Transient localized states. A selection of eigenstates (red color scheme) near the Fermi level is shown for the considered prototype materials at four temperatures, where the gray scale represents the corresponding frozen deformation potential. An increasing temperature leads to more spatially confined states that are linked to fugitive, Anderson-localized states in transient dynamics.}\label{Fig:quasibound_states}
\end{figure}

\begin{figure*}[t!]
\centering
\includegraphics[width=0.98\textwidth]{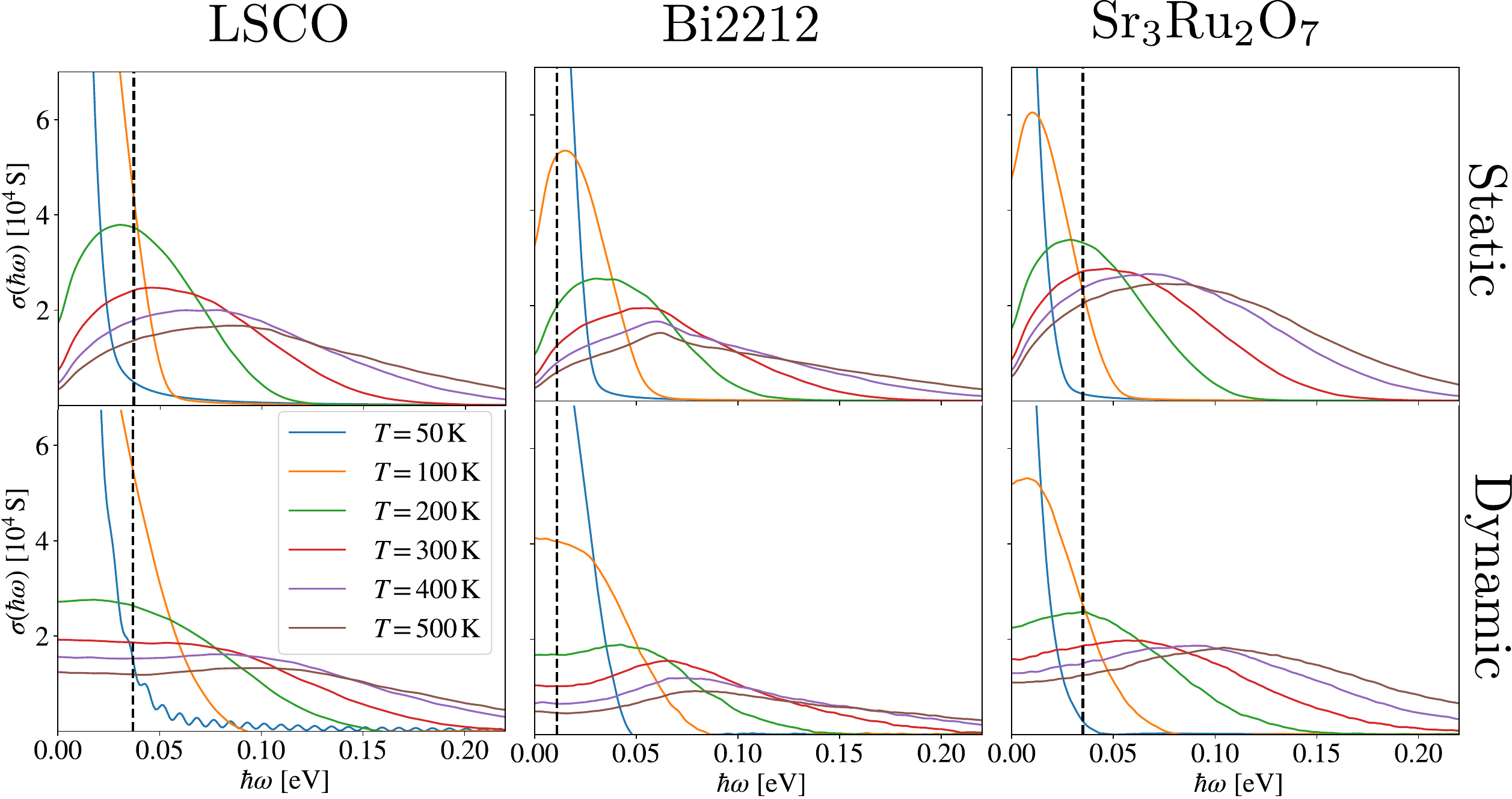}
\caption{Frozen versus dynamical quantum acoustic vibration field. The upper and lower panels display the optical conductivity for the three materials at different temperatures, resolved within the static and dynamical potential landscapes averaged over 100 and 10 realizations, respectively. The back dash line marks the Debye frequency of the given material below which the frozen potential assumption breaks down. A key dynamical effect is the saturation of conductivity in the regime $\omega \lesssim \omega_{D}$, instead of the suppression evident in the static case. However, regardless of the deformation potential dynamics and the chosen material, the optical conductivity peak shifts from the Drude-peak edict of situating at $\omega = 0$ to higher energies and broadens as the temperature increases.}
\label{Fig:frozen_vs_dynamical}
\end{figure*}

It appears that, regarding photoabsorption,  electrons are insensitive to the lattice dynamics when $\omega \gg w_{D}$. In other words, the deformation landscape appears as if it is stationary for an electron in this regime, a notion confirmed by our results below. 

Motivated by this, we initially freeze the potential and study its transitory electronic eigenstates.  Since the allowed transitions take place near the Fermi level, we are allowed to focus on the states lying within the stack of $ \varepsilon_{F} \pm 3 k_{\textrm{B}} T$. Fig~\ref{Fig:quasibound_states} shows examples of eigenstates for the three materials at different temperatures along with the profile of the deformation potential. Despite the diversity among the physical settings, all three materials share similar qualitative characteristics: Instead of being spatially extended, as observed at lower temperatures ($E_{F}/V_{\textrm{rms}} \gtrsim 1$), the relevant states appear to be localized in the dips of the potential at temperatures $E_{F}/V_{\textrm{rms}} \lesssim 1$ where the potential fluctuation $V_{\textrm{rms}}$ is comparable to Fermi energy $E_{F}$. These eigenstates of the frozen potential relate to the localized states associated with the transient dynamics.

Next, we compute the corresponding optical conductivity $\sigma(\hbar \omega)$ within the Kubo formalism first by numerically diagonalizing the Hamiltonian with the potential given
in Eq.~\ref{eq:VDcl}.~\footnote{See Appendix~\ref{Appendix:frozen_approximation} for a detailed description about the Kubo calculations carried out in the case of a frozen potential. In the nutshell, we first numerically determine the eigenstates of an ensemble of the frozen deformation potential, and then employ them to compute the optical conductivity via the Kubo formula.} In the upper panel of Fig~\ref{Fig:frozen_vs_dynamical}, we present the optical conductivity at various temperatures for the three chosen materials, averaged over an ensemble of 100 random realizations of the deformation potential. With increasing temperature, the optical conductivity evolves from the Drude-peak behavior of having a sharp maximum value located at $\omega = 0$ into a displaced peak: the maximum conductivity point steadily shifts towards higher energies $\hbar \omega$ and the conductivity peak profile broadens. 

\begin{figure}[t]
\centering
\includegraphics[width=0.48\textwidth]{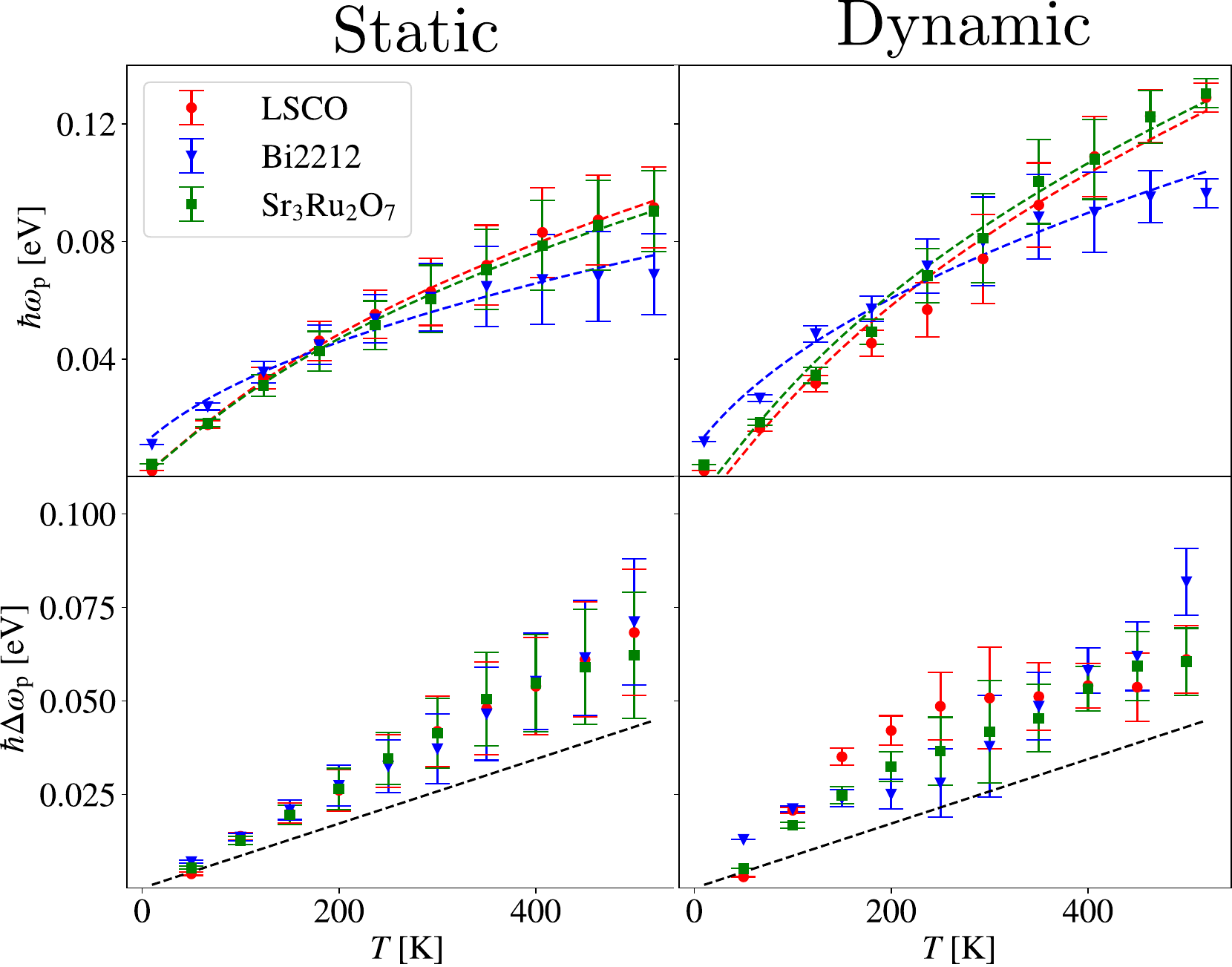}
\caption{Position and width of the Drude peak. The panels show the mean location (top) and width (bottom) of the optical conductivity peak as a function of temperature for the studied materials, averaged over 100 (static) and 10 (dynamic) deformation potential realizations, with error bars representing the standard deviation within the given ensemble. Specifically, as the temperature increases, there is a generic upward shift in the Drude peaks towards higher energies (upper panel), accompanied by a broadening of the peaks (lower panel). Colored dashed curves show the fittings for the DDP location estimated with the assumption that it is mostly determined by the strength of the deformation potential ($\hbar\omega_p \propto V_{\textrm{rms}}$). Furthermore, the width of the DDPs live near to Planckian bound ($\hbar \Delta \omega_p \sim k_BT$), which is indicated by the black dashed line.}
\label{Fig:position_and_width}
\end{figure}

In addition, we show the temperature dependence of the peak locations and their width in Fig.~\ref{Fig:position_and_width}. We determine the peak location $\hbar \omega_{\textrm{p}}$ as the energy at which the optical conductivity $\sigma(\hbar \omega)$ reaches its maximum. The peak width $ \hbar \Delta \omega_{\textrm{p}}$ is then defined in a similar manner as in Ref.~\cite{Scipost_3_025_2017}: the distance between the maximum and the optical conductivity point in the high-energy tail where the height of the maximum is dropped by 50\%. Fig.~\ref{Fig:position_and_width} further confirms and quantifies the migration and broadening of the DDP with increasing temperature present in  Fig.~\ref{Fig:frozen_vs_dynamical}. 

%\subsection{Dynamical field}
 
To further validate the frozen potential results above, we expand our DDP analysis by computing the optical conductivity while considering the temporal evolution of the deformation potential. Nonetheless, we can still determine the conductivity $\sigma(\hbar \omega)$ by utilizing the Kubo formalism.\footnote{See Appendix~\ref{Appendix:dynamical_field} explicating the computation of the optical conductivity within the Kubo formulation in the case of the dynamical deformation potential. In summary, we take the already resolved eigenstates, and let them evolve under time-depend deformation potential defined in Eq.~\ref{eq:VDcl}. By computing the velocity autocorrelation for all time steps, we numerically determine the optical conductivity in the case of a dynamical lattice disorder field within the Kubo formulation.} In short, we take advantage of the already defined eigenstates of the frozen deformation potentials as initial conditions, and let it thereafter unfold according to Eq.~\ref{eq:VDcl}. 

The lower panel of Fig.~\ref{Fig:frozen_vs_dynamical} shows the optical conductivity spectrum of the three materials at different temperatures in the case of a dynamical potential field, averaged over 10 realizations of the distorted potential landscape. As expected, the dynamical conductivity spectrum deviates from the frozen potential prediction when $\omega \lesssim \omega_{D}$ is indicated by the black dash line in Fig.~\ref{Fig:frozen_vs_dynamical}. Instead of strong suppression near to the d.c. conductivity as in the static landscape situation, we observe a saturation of the optical conductivity within an energy window of the order of $0.1\, \textrm{eV}$ near the zero frequency. This can be interpreted as the reversed adiabatic approximation where the external electric field of frequency $\omega\ll\omega_D$ is a slow varying degree of freedom and thus is roughly static compared to the fluctuations of the lattice, yielding virtually the same conductivity as the d.c. conductivity and manifesting as a conductivity plateau below the Debye frequency. Nevertheless, there is still a generic trend similar to the frozen potential approximation: the higher temperature yields a more substantial DDP, suggesting that the transiently localized states are at play in both cases. This tendency is also evident in Fig.~\ref{Fig:position_and_width} where the increase in temperature moves the DDP to higher absorption frequencies while broadening the peak at the same time.     

%\section{Discussion and summary}

The physical picture behind the observed DDP evolution is that the increase in temperature has a two-fold effect. First, it yields stronger spatially localized, transient electronic states at the Fermi energy (shift to higher frequencies); electrons either residing in local potential wells (frozen) or nesting in instantaneous potential pockets (dynamic). These local wells or nests become more energetically confining as the deformation potential strengthens with the rising temperature. As a result, the location of the DDP roughly scales like $\hbar\omega_{\textrm{p}} \sim V_{\textrm{rms}}$, which defines the fitting 
illustrated by the colored dashed curves in the upper panel of Fig.~\ref{Fig:position_and_width}.~\footnote{More specifically, we define the peak location of a displaced Drude peak as $\hbar \omega_{\textrm{p}} = a + b V_{\textrm{rms}}$, where $a$ and $b$ are fitting parameters, and the temperature dependence is included in the root mean square of the deformation potential $V_{\textrm{rms}}$ that is explicitly presented in Appendix~\ref{Appendix:def.potential}}
In general, the peak location migrates like $\hbar\omega_{\textrm{p}} \sim (k_B T)^{3/2}$ at low temperatures $T \ll T_D$ and  $\hbar\omega_{\textrm{p}} \sim (k_B T)^{1/2}$ at high temperatures $T \gg T_D$ . In addition, the presence of eigenstate localization caused by the frozen lattice disorder is known to result in  band tails in the density of states~\cite{Heller22} that has later shown to persist even under quantum-acoustical lattice dynamics~\cite{Phys.Rev.B_107_224311_2023}. 

On the other hand, a higher temperature permits a wider energy window for electronic transitions to occur (the broadening of the peak). Whereas the transition element between the (momentarily) localized states dictates the location of DDP, the width is instead determined by the broadening function characterizing the energetically allowed transition. As indicated by the black dashed line in the lower panel of Fig.~\ref{Fig:position_and_width}, the widths of the DDP behave roughly as $\hbar\Delta  \omega_{\textrm{p}} \sim k_{\textrm{B}}T$ that is interestingly more accurate in the case of the dynamical potential landscape. The quantum-acoustical DDP is thus intimately connected to the ambiguous Planckian timescale $\hbar/k_{\textrm{B}}T$ that underpins the linear-in-temperature resistivity exhibited by numerous families of bad and strange metals (for comparison, see Ref.~\cite{Scipost_3_025_2017}). In addition, this observed correlation between the width of the DDP and Planckian behavior supports the prospect of the near-universal transport by transient dynamics reported in Ref.~\cite{Heller22,nature_manuscript}.

In the quantum-acoustic DDP scheme, there are no extrinsic sources, such as defects or impurities, which could also generate or enhance a shift in optical conductivity (see, e.g., Refs.~\cite{Phys.Rev.Lett_81_2132_1998, Phys.Rev.B_57_R11081_1998, Phys.Rev.B_49_12165_1994}). In other words, the disorder at the origin of our DDP is \emph{self-generated}, arising from the existence of thermally fluctuating lattice degrees of freedom that significantly affect the charge carrier dynamics. In particular, our DDP gives a unique temperature-dependent fingerprint, clearly distinguishing it from an impurity-induced DDP. The general trend of a DDP seen in Fig.~\ref{Fig:frozen_vs_dynamical} qualitatively agrees with experimental observations~\footnote{There is a growing multitude of experimental evidence on the emergence of a displaced Drude peak in several strange metals that that widens and moves toward higher frequencies with increasing temperature, see, e.g., K. Takenaka, et al., Phys. Rev. B 65, 092405 (2002); A. F. Santander-Syro, et al., 
Phys. Rev. Lett. 88, 097005 (2002); C. C. Homes, et al. Phys. Rev. B 74, 214515 (2006); J. P. Falck, et al., Phys. Rev. B 48, 4043 (1993); J. P. Falck, et al., Phys. Rev. B 48, 4043 (2004).
}, also supporting the scaling behavior of the quantum-acoustical DDP illustrated in Fig.~\ref{Fig:position_and_width}.~\footnote{The experimental datasets collected in Refs.~\cite{scipost.phys_11_39_2021, Scipost_3_025_2017} reveals a prominent role of thermal fluctuations as expected within the present framework. As reported in Ref.~\cite{scipost.phys_11_39_2021}, the temperature dependence of the DDP peak in various material follows $\hbar \omega_p \propto T^{\alpha}$, where $0.5 \le \alpha \le 1.5$. Moreover, 
in various classes of strange/bad metals, the width of the peak is experimentally found to be lie at the Planckian bound ($\hbar \Delta \omega_p \approx k_B T$), as collectively presented in Ref.~\cite{Scipost_3_025_2017}. Both of these experimental features are in agreement of the theory of the quantum-acoustical DDP.} Noteworthy, the transient dynamics driving the birth of an acoustical DPP resides within a dynamical regime of nonperturbative and coherent electron-lattice motion, thus lying outside the reach of the conventional perturbative or Boltzmann transport methods (see Ref.~\cite{Heller22}). In fact, many bad and strange metals are on the verge of a transient localization and/or have strong electron-phonon coupling~\cite{Lanzara2001,rev.mod.phys_92_031001_2020}.

Moreover, alongside the DDP formation and Planckian resistivity, the deformation potential perspective offers a natural pathway for charge carriers in strange metals to cross the MIR bound with impunity at high temperatures~\cite{Philos.Mag_84_2847_2004, Rev.Mod.Phys_75_1085_2003}, as already asserted in Ref.~\cite{Heller22}. We explicitly demonstrated the violation of MIR limit with quantum acoustics in another paper~\cite{nature_manuscript}. Our approach also carries the potential to enlighten the perplexing phenomenon of pseudogaps~~\cite{science_314_1888_2006, rep.prog.phys_62_61_1999, Rev.Mod.Phys_75_473_2003, Rev.Mod.Phys_77_721_2005, Rev.Mod.Phys_79_353_2007} and charge density waves~~\cite{Rev.Mod.Phys_60_1129_1988, nature_477_191_2011, nat.phys_8_871_2012, science_337_6096_2012}. In particular, we see that when electron motion strongly couples and synchronizes with low-energy lattice vibration modes, it creates a favorable environment for incommensurate charge density order. Likewise, a similar resonance promoted by the deformation potential could result in temperature-dependent pseudogaps, i.e., a substantial suppression in the density of the low-energy excitations, which eventually melt away leaving the pseudogap phase regime. We aim to study these considerations in future research.

In conclusion, we have introduced the phenomenon of quantum-acoustical Drude peak displacement, which involves the temperature-dependent shift and broadening of the optical conductivity peak to finite frequencies, demonstrated here for three archetypal strange metals. Overall, the coherent state picture of lattice vibrations, which has always been at one's disposal but not utilized, provides a fresh perspective on the investigation of the mysteries of bad and strange metals. The manifested shift in perspective simply comes from the coherent state limit of quantum acoustics.

\begin{acknowledgments}
We are thankful for the useful discussions with D. Kim, B. Halperin, S. Das Sarma, J. H. Miller Jr, R. L. Greene, R. Lobo and A. P. Mackenzie. Furthermore, J.K.-R. thanks the Emil Aaltonen Foundation, Vaisala Foundation, and the Oskar Huttunen Foundation, and A. M. G. thanks the Harvard Quantum Initiative for financial support.
\end{acknowledgments}

\appendix

\section{Deformation potential}\label{Appendix:def.potential}

Here we provide a brief overview of the coherent state formalism and the origin of the deformation potential; for a comprehensive discussion about the topic, see Ref.~\cite{Heller22}. 

In general, a lattice deformation is treated in terms of a quantum longitudinal displacement field $\hat{\mathbf{u}}(\mathbf{r},t)$. The
longitudinal displacement of an atom at a position $\mathbf{r}$ from its equilibrium position is
\begin{eqnarray*}\label{displament_field}
\begin{aligned}
    \hat{\mathbf{u}}(\mathbf{r},t)
    &=
    -i\sum_{\mathbf{q}}\hat{\mathbf{q}}\sqrt{\frac{\hbar}{2\rho \mathcal{V}\omega_{\mathbf{q}}}}
    \\
    &\qquad\qquad\quad\times
    \left( a_{\mathbf{q}}e^{-i\omega_{\mathbf{q}}t}
    -
    a_{-\mathbf{q}}^\dag e^{i\omega_{\mathbf{q}}t} \right)
    e^{i\mathbf{q}\cdot\mathbf{r}},
    \label{eq:uquantum}
\end{aligned}
\end{eqnarray*}
where $\mathbf{q}$ is wave vector of a normal mode, $\omega_{\mathbf{q}}$ is phonon frequency, $\rho$ is mass density of the lattice, $\mathcal{V}$ is volume (area in 2D), and $t$ is time. We omit writing the polarization vector and phonon branch indices in the subscripts since we will only deal with the longitudinal acoustic modes of the lattice vibrations, the main scattering mechanism for the charge carriers. Subsequently, we define the deformation potential as the first-order correction in the expansion of band energy due to atomic displacements in the following way
\begin{equation*}
    \begin{split}
        \hat{V}_D(\mathbf{r},t) = & E_d\nabla\cdot\hat{\mathbf{u}}(\mathbf{r},t)\\
        = & \sum_\textbf{q}^{|\mathbf{q}|\le q_D}E_d\sqrt{\frac{\hbar}{2\rho\mathcal{V}\omega_\mathbf{q}}}|\mathbf{q}|(a_\mathbf{q}+a_\mathbf{-q}^\dagger)e^{i\textbf{q}\cdot\textbf{r}}\\
        = & \sum_\textbf{q}^{|\mathbf{q}|\le q_D}g_\mathbf{q}(a_\mathbf{q}+a_\mathbf{-q}^\dagger)e^{i\mathbf{q}\cdot\mathbf{r}},
    \end{split}
\end{equation*} 
where the deformation potential constant $E_d$ characterizes the coupling between electrons and the lattice, and the mode summation is restricted by the Debye wavenumber $q_{D}$. For convenience, we convert the material constant into the mode-depend co-factor of
\begin{equation*}
    g_{\mathbf{q}} = E_d\sqrt{\frac{\hbar}{2\rho\mathcal{V}\omega_\mathbf{q}}}|\mathbf{q}| = E_d\sqrt{\frac{2 \hbar \vert \mathbf{q} \vert}{\rho \mathcal{V} v_s}},
\end{equation*}
where the latter form is achieved by assuming the linear dispersion $\omega_{\mathbf{q}} = v_s \vert \mathbf{q} \vert$, coupling given by the speed of sound $v_s$. Thus, the electron-phonon interaction term corresponds to the Fröhlich Hamiltonian in the main text.

The next ingredient we introduce is the coherent state picture. Within this framework, each normal mode of lattice vibration with a wave vector $\mathbf{q}$ is associated with a coherent state $\vert \alpha_{\mathbf{q}} \rangle$. By employing the independence of normal modes, entire lattice vibrations can be expressed as the product state of the coherent states $\vert\alpha_{\mathbf{q}}\rangle$ as a single multimode coherent state of
\begin{eqnarray*}
\begin{aligned}
\vert \bm{\alpha} \rangle = \prod_{\mathbf{q}}\vert\alpha_{\mathbf{q}} \rangle,
\end{aligned}
\end{eqnarray*}
whose expectation value defines the deformation potential
\begin{eqnarray*}
\begin{aligned}
    V_D(\mathbf{r},t)
    &= \langle \bm{\alpha} \vert \hat{V}_D(\mathbf{r},t) \vert \bm{\alpha} \rangle
    \\
    &=
    \sum_{\mathbf{q}}^{\vert \mathbf{q} \vert \le q_D}
    g_{\mathbf{q}}
    (\alpha_{\mathbf{q}}e^{-i\omega_{\mathbf{q}}t}
    +
    \alpha_{-\mathbf{q}}^* e^{i\omega_{\mathbf{q}}t})
    e^{i\mathbf{q}\cdot\mathbf{r}}.
\end{aligned}
\end{eqnarray*}
This quasi-classical field of lattice vibrations corresponds to the deformation potential presented in the main text by considering a thermal coherent state 
\begin{equation*}
    \alpha_\mathbf{q}=\sqrt{\langle n_{\mathbf{q}}\rangle_{\textrm{th}}}\exp(i\varphi_\mathbf{q}),
\end{equation*}
where $\sqrt{\langle n_{\mathbf{q}}\rangle_{\textrm{th}}}=\vert\alpha_\mathbf{q}\vert$ is the thermal amplitude given by the Bose-Einstein distribution and $\varphi_{\mathbf{q}}=\arg(\alpha_{\mathbf{q}})$ is the random phase of the coherent state determining the initial conditions. 

The deformation potential in itself is a curious mathematical object. First, it is homogeneously random in space and in time, meaning the probability distribution of deformation potential is independent on a position $\mathbf{r}$ or time t given that the phases $\varphi_{\mathbf{q}}$ are random variables. Thus, each reasonably large spatiotemporal section of the potential is statistically indistinguishable from another. Second, although the deformation potential averages to zero, its root-mean-square identifies the strength of lattice disorder, growing in temperature as
\begin{equation*}
 V_{\textrm{rms}}^{2} = \frac{2 E_d^2 \hbar}{\pi \rho v_s} \int_0^{q_D} \frac{q^2\textrm{d} q}{e^{\hbar v_s q/k_BT}-1}.
\end{equation*}
In particular, we employ the expression above to analyze the evolution of the peak location of a displaced Drude peak as a function of temperature. 

Regarding electron dynamics, we focus on the quantum dynamics of an electron under the following time-dependent Hamiltonian:
\begin{eqnarray*}
    \mathcal{H}_0 = \frac{\vert \mathbf{p} \vert^2}{2 m} + V_{D}(\mathbf{r}, t),
\end{eqnarray*}
where $m$ is the effective (band) mass of the electron. This effective Hamiltonian is the electron part of the considered Fröhlich Hamiltonian described within the effective mass approximation. 

A comparison of electrons lying in the neighborhood of the Fermi energy with the strength of the deformation potential determines whether the scattering can be treated perturbatively or not~\cite{nature_manuscript}. Furthermore, quantum
coherence of electrons is important in scattering when the wavelength of the Fermi wavelength $\lambda_F$ is not much less than twice the lattice constant $a$. It should be emphasized that the term ``coherence'' is employed here to describe the spatial phase coherence of the electronic wavefunction, and should not be conflated with the ``coherent-versus-incoherent-metals'' nomenclature, which pertains to the breakdown of the quasi-particle paradigm. In summary, dynamics is coarsely classified as
\begin{equation*}
\frac{E_F}{V_{\textrm{rms}}} =
\begin{cases}
	\bar{E}\gg 1  & \rightarrow \;  \text{Perturbative} \\
	\bar{E}\sim 1 \; \text{or} \; \bar{E}\ll 1  & \rightarrow \;  \text{Nonperturbative},
\end{cases}
\end{equation*}
and
\begin{equation*}
\frac{\lambda_F}{2a} =
\begin{cases}
	\bar{\lambda}\ll 1  & \rightarrow \;  \text{Incoherent} \\
	\bar{\lambda}\sim 1 \; \text{or} \; \bar{\lambda}>1  & \rightarrow \;  \text{Coherent}.
\end{cases}
\end{equation*}

\section{Material parameters}\label{Appendix:material_parameters}

Table~\ref{Table:material_parameters} contains the relevant material parameters for determining the optical conductivity within the deformation potential scheme. The strange metals in the Table below possess two characteristic attributes: relatively high deformation potential constant and low Fermi energy compared to normal metals.
However, we want to point out the robustness of the results reported in the main text against the reasonable range of material parameters, in a similar manner as analyzed in Ref.~\cite{Aydin}. In other words, reasonable deviations from the given values do not change the qualitative conclusions of the work. 

\begin{table}[h!]
\begin{tabular}{c|ccc}
 & LSCO & Bi2212 & Sr\textsubscript{3}Ru\textsubscript{2}O\textsubscript{7}\\
\hline \hline\\
$n$ [$10^{27}\, \textrm{m}^{-3}$] & 7.8 & 6.8 & 0.5\\
\\
$m^*$ [$(m_e)$] & 9.8 & 8.4 & 6.8\\
\\
$v_s$ [m/s] & 6000 & 2460 & 5850\\
\\
$E_d$ [eV] & 20 & 10 & 20\\
\\
$\rho$ [$10^{-6}\,\text{kg/m}^2$] & 3.6 & 5.2 & 8.9\\
\\
$E_F$ [eV] & 0.12 & 0.15 & 0.03\\
\\
$a$ [Å] & 3.8 & 5.4 & 3.9\\
\\
$T_D$ [K] & 427 & 123 & 406 \\
\\
\end{tabular}
\caption{Material parameters that are used for three different strange metals.}\label{Table:material_parameters}
\label{table}
\end{table}

\section{Optical conductivity}

Here we discuss the linear response theory behind our optical conductivity results. In general,  
the (complex) conductivity tensor $\sigma$ relates the current density $j_{x}(t)$ to the applied electric field $\textrm{Re}[E_{x} \exp(i\omega t)]$.
Within the Kubo formalism, if the system is in thermal equilibrium with a heat reservoir with temperature $T$, the conductivity tensor for non-interacting particles is 
\begin{equation*}\label{eq:Kubo_origin}
    \sigma(\omega) = \lim_{\eta\to 0^+}\frac{\mathcal{V}}{\hbar \omega} \int_{0}^{\infty} \textrm{d}t\, \Big\langle [\hat{j}_{x}(t), \hat{j}_{x}(0)] \Big\rangle e^{i(\omega  + i \eta)t}+\frac{ie^2n}{m\omega},
\end{equation*}
 $\mathcal{V}$ is the volume of the system, $m$ is the effective mass of the electron, $n$ represents the number density of electrons and $\langle \cdots \rangle$ refers to the averaging over equilibrium thermal ensemble. We also include an infinitesimal increment $i \eta$ to ensure that the integrand vanishes exponentially as $t \rightarrow \infty$, and thus the integral is well-defined.

\subsection{Frozen approximation (Adiabatic limit)}\label{Appendix:frozen_approximation}
 
In the regime of $\omega\gg\omega_D$, the time scale of the perturbing electric field on the system is much smaller than the timescale describing the motion of lattice. Therefore, a good, first approximation would be to assume that lattice is motionless. In fact, this frozen lattice approximation relates to the adiabatic approximation which is basically omitting the dynamics of the slow degrees of freedom when considering the fast degrees of motion.

When we freeze the deformation potential, we first solve the electronic states $\vert n \rangle$ and corresponding energies $\varepsilon_n$ determined by the equation
\begin{equation*}
    \left( \frac{\mathbf{p}^2}{2m^*} +  V_D(\mathbf{r}) \right)\vert n \rangle  = \varepsilon_n \vert n \rangle.
\end{equation*}
Then, by utilizing this eigenbasis, the Kubo formula for the optical conductivity takes the form of 
\begin{equation*}\label{eq:Kubo_static}
\begin{split}
\sigma(\omega) = -2 &\mathrm{Re} \Bigg[\lim_{\eta\to0^+}\frac{e^2 \hbar}{\mathcal{V}} \sum_{n,m} \frac{f(\varepsilon_n) - f(\varepsilon_m)}{\varepsilon_n - \varepsilon_m} \times\\ &\frac{\vert \langle n \vert\hat{v}_x\vert m \rangle \vert^2}{(\hbar\omega+\varepsilon_n - \varepsilon_m)^2+\eta^2}\eta \Bigg],
\end{split}
\end{equation*}
where the factor of two represents the spin degree of freedom. 

\begin{figure*}[t!]
\centering
\includegraphics[width=0.7\textwidth]{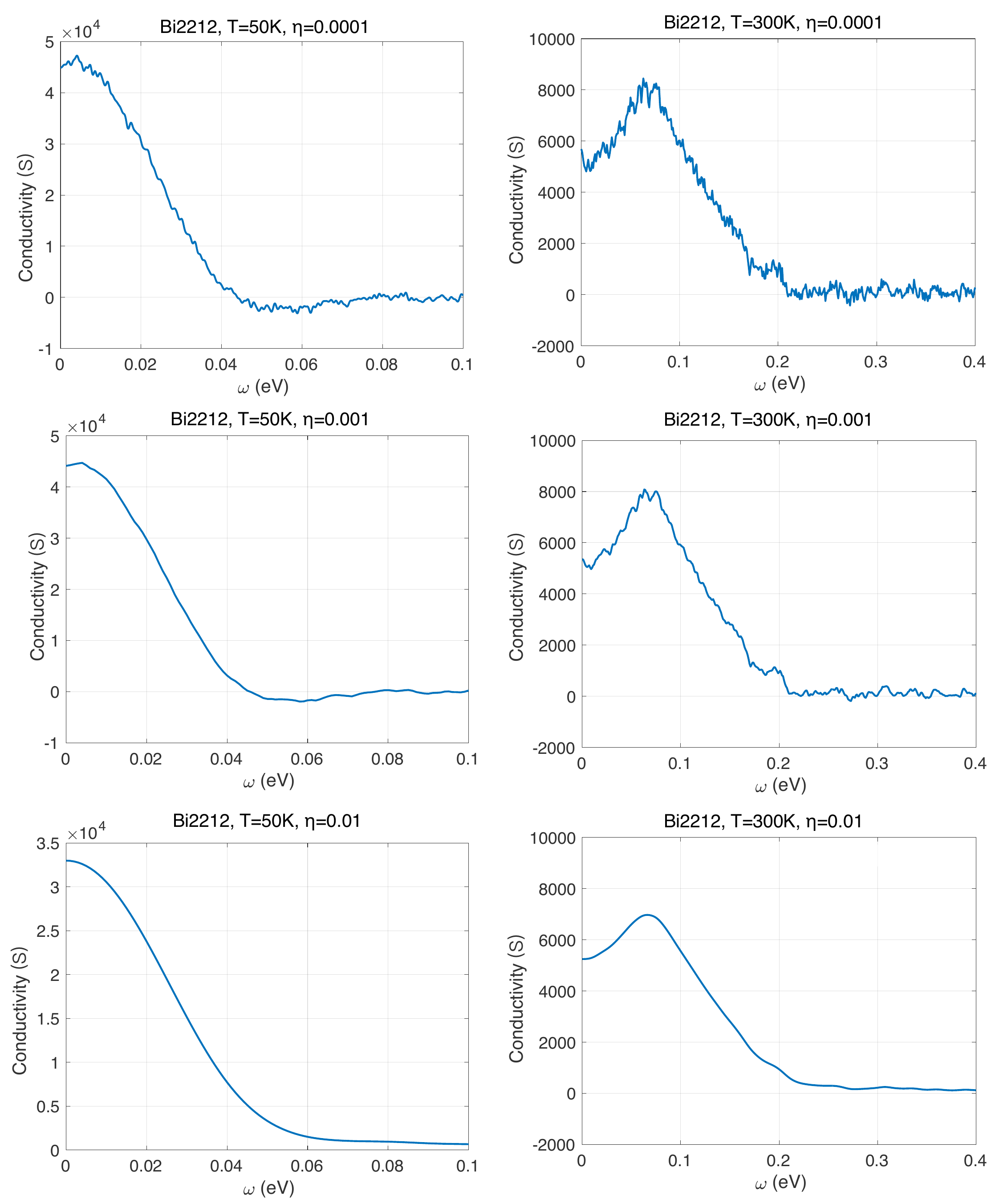}
\caption{Figure shows the optical conductivity for Bi2212 in the case of three different $\eta$ parameters at two temperatures that were computed for the dynamical field, i.e., in the diabatic limit. At both temperatures of $50\, \textrm{K}$ and $300\, \textrm{K}$, the increase of the parameter $\eta$ leads to an artificial smoothing of the optical conductivity features. Nevertheless, even a large value of $eta = 0.01$ compared to energy level spacing does not erase the general trend of the conductivity, only washing out the small scale fluctuations which are more prominent at higher temperatures due to a stronger deformation potential.}
\label{Fig:eta_parameter}
\end{figure*}

The artifact parameter $\eta$ should be theoretically infinitely small, but in practice, we choose the parameter $\eta$ to be small enough not to impact the results. In Fig.~\ref{Fig:eta_parameter}, we demonstrate that this parameter has an effect of smoothing the optical conductivity fluctuations which are more eminent. Nonetheless, the general behavior survives even under a poor choice of the parameter $\eta$ where it is larger than the average energy spacing between individual eigenenergies.  

In our simulations, the value of $\eta$ is first assumed to roughly be the order of the energy gap between adjacent energy levels, thus guaranteeing that the fluctuations of the optical conductivity will not be too large. Second, we demand that the broadening of energy levels caused by $\eta$ is smaller than the thermal smearing ($\sim k_BT$) arising from the Fermi function in the Kubo formula above. Subsequently, the effect of a finite parameter $\eta$ is negligible in the present optical conductivity results.

\subsection{Dynamical field (Diabatic limit)}\label{Appendix:dynamical_field}

On the other hand, when $\omega\lesssim\omega_D$, the frozen potential approximation is not valid. However, we can still determine the optical conductivity by employing the Kubo formalism in the following manner,
\begin{equation*}\label{eq:Kubo_dynamic}
    \begin{split}
    \sigma(\omega)  = -2 &\mathrm{Re} \Bigg[ \lim_{\eta\to 0^+}\frac{e^2}{\mathcal{V}} \sum_{n,m} \frac{f(\varepsilon_n) - f(\varepsilon_m)}{\varepsilon_n - \varepsilon_m}\times\\
    &\int_0^\infty\mathrm{d}t\,\langle n \vert\hat{v}_x(t)\vert m \rangle\langle m\vert\hat{v}_x\vert n \rangle e^{i(\omega+i\eta)t} \bigg],
    \end{split}
\end{equation*}
where $\hat{v}_x(t)$ is the velocity operator in the Heisenberg picture, and all the eigenstates and eigenvalues are taken to be that of the initial condition. This dynamical formulation reduces to the Kubo formula above when the deformation potential is treated as static, but it is also well-defined for dynamical deformation potential.

In the Kubo formula above, the time integration is numerically carried out up to a finite time $T$ that is decided by $\eta$. In practice, we first choose the parameter $\eta$ with the same criteria as in the frozen potential case, and then the maximum propagation time is set by $\eta T \ge 5$. We estimate the integration error to be 
\begin{equation*}
\begin{split}
    \Delta(T) &= \Bigg \vert \int_0^\infty\mathrm{d}t f(t)\mathrm{e}^{\mathrm{i}(\omega+\mathrm{i}\eta)t}-\int_0^T\mathrm{d}t f(t)\mathrm{e}^{\mathrm{i}(\omega+\mathrm{i}\eta)t} \Bigg \vert\\ &\le \frac{M T}{\exp\left( \eta T \right) - 1},
    \end{split}
\end{equation*}
where $f(t)$ is the velocity autocorrelation and $M$ is a constant. This truncation causes an error in an slack of $~\hbar/T$ in the vicinity of the d.c. conductivity $\sigma(0)$, but the condition of $\eta T \ge 5$ ensures that the computed optical conductivity is numerically robust.

\bibliography{references}% Produces the bibliography via BibTeX.

\end{document}